\documentclass[]{aa}
\usepackage{graphics,epsfig,txfonts}
\usepackage{natbib}
\bibpunct{(}{)}{;}{a}{}{,}

\newcommand{\ergcm}[1]{$\times 10^{#1}$ erg cm$^{-2}$ s$^{-1}$}

\newcommand{\ergs}[1]{$\times 10^{#1}$ erg s$^{-1}$}

\newcommand{\hcm}[1]{$\times 10^{#1}$ cm$^{-2}$}
\newcommand{\ohcm}[1]{$10^{#1}$ cm$^{-2}$}
\newcommand{\expo}[1]{$\times 10^{#1}$}

\newcommand{\kms}{km s$^{-1}$}
\newcommand{\nh}{N$_{\rm H}$}

\newcommand{\SII}{[\ion{S}{ii}]}
\newcommand{\OIII}{[\ion{O}{iii}]}
\newcommand{\Halpha}{H${\alpha}$}
\newcommand{\ltsima}{$\buildrel < \over \sim$}
\newcommand{\lsim}{\lower.5ex\hbox{\ltsima}}
\newcommand{\gtsima}{$\buildrel > \over \sim$}
\newcommand{\gsim}{\lower.5ex\hbox{\gtsima}}
\newcommand{\msun}{M$_{\odot}$}

\newcommand{\rahour}{\hbox{\ensuremath{^{\rm h}}}}
\newcommand{\ramin}{\hbox{\ensuremath{^{\rm m}}}}

\newcommand{\xmm}{{\it XMM-Newton}}
\newcommand{\cxo}{{\it Chandra}}
\newcommand{\einstein}{{\it Einstein}}
\newcommand{\asca}{ASCA}
\newcommand{\ROSAT}{ROSAT}
\def\SNR{\mbox{{SNR\,J0453--6829}}}
\def\p0{\phantom{0}}

\begin{document}
 
\title{Multi-frequency observations of \SNR\ in the LMC}
\subtitle{A composite supernova remnant with a pulsar wind nebula}

\author{F.~Haberl\inst{1} \and M.\,D.~Filipovi{\'c}\inst{2} \and L.\,M.~Bozzetto\inst{2} \and E.\,J.~Crawford\inst{2} 
        \and S. D. Points \inst{3} \and W.~Pietsch\inst{1} 
        \and A.\,Y.~De~Horta\inst{2} \and N.~Tothill\inst{2} \and J.\,L.~Payne\inst{2} \and M.~Sasaki\inst{4}}

\titlerunning{Multi-frequency observations of \SNR}
\authorrunning{Haberl et al.}
 
\institute{Max-Planck-Institut f\"ur extraterrestrische Physik,
           Giessenbachstra{\ss}e, 85748 Garching, Germany, \email{fwh@mpe.mpg.de}
	   \and
           University of Western Sydney, Locked Bag 1797, Penrith South DC, NSW1797, Australia
	   \and
           Cerro Tololo Inter-American Observatory, National Optical
           Astronomy Observatory, Cassilla 603 La Serena, Chile 
           \and
           Institut f{\"u}r Astronomie und Astrophysik T{\"u}bingen, Sand 1, D-72076 T{\"u}bingen, Germany
	   }
 
\date{Received 6 February 2012 / Accepted 21 June 2012 }

\abstract
{The Large Magellanic Cloud (LMC) is rich in supernova remnants (SNRs) which can be investigated in detail with radio, optical and X-ray observations. \SNR\ is an X-ray and radio-bright remnant in the LMC, within which previous studies revealed the presence of a pulsar wind nebula (PWN), making it one of the most interesting SNRs in the Local Group of galaxies.}
{We study the emission of \SNR\ to improve our understanding of its morphology, spectrum, and thus the emission mechanisms in the shell and the PWN of the remnant.}
{We obtained new radio data with the Australia Telescope Compact Array and analysed archival \xmm\ observations of \SNR. We studied the morphology of \SNR\ from radio, optical and X-ray images and investigated the energy spectra in the different parts of the remnant.}
{Our radio results confirm that this LMC SNR hosts a typical PWN. The prominent central core of the PWN exhibits a radio spectral index $\alpha_{\rm Core}$ of --0.04$\pm$0.04, while in the rest of the SNR shell the spectral slope is somewhat steeper with $\alpha_{\rm Shell}$ = --0.43$\pm$0.01. We detect regions with a mean polarisation of $P\cong$ (12$\pm$4)\% at 6~cm and (9$\pm$2)\% at 3~cm. The full remnant is of roughly circular shape with dimensions of (31$\pm$1)~pc $\times$ (29$\pm$1)~pc. The spectral analysis of the \xmm\ EPIC and RGS spectra allowed us to derive physical parameters for the SNR. Somewhat depending on the spectral model, we obtain for the remnant a shock temperature of around 0.2~keV and estimate the dynamical age to 12\,000-15\,000 years. Using a Sedov model we further derive an electron density in the X-ray emitting material of 1.56\,cm$^{-3}$, typical for LMC remnants, a large swept-up mass of 830~\msun, and an explosion energy of 7.6$\times 10^{50}$~erg. These parameters indicate a well evolved SNR with an X-ray spectrum dominated by emission from the swept-up material.}
{}


\maketitle

\section{Introduction}
\label{intro}

The Magellanic Clouds (MCs) are considered to be one of the most ideal environments when it comes to the investigation of various supernova remnant (SNR) types and their different evolutionary stages. While their relatively small distance is very favourable for detailed studies, the MCs are also located in one of the coldest parts of the radio sky, allowing us to detect and investigate radio emission with little disturbing Galactic foreground radiation \citep{1991A&A...252..475H}. As they are located outside of the Galactic plane, the influence of dust, gas and stars is small, reducing the absorption of soft X-rays. Particularly, the Large Magellanic Cloud (LMC) at a distance of 50~kpc \citep{2008MNRAS.390.1762D}, allows for detailed analysis of the energetics of various types of remnants. In the radio-continuum regime SNRs are well-known for their strong and predominantly non-thermal radio emission, which is characterised by a typical spectral index of $\alpha\sim-0.5$ (as defined by $S\propto\nu^\alpha$). However, this value shows a significant scatter due to the wide variety of SNR types and this variance can be used as an age indicator for the SNR \citep{1998A&AS..130..421F}.

SNRs have a significant influence on the structure of the interstellar medium (ISM). The appearance of spherically symmetric shell-like structures is very often perturbed by interaction with a non-homogeneous structure of the ISM. SNRs influence the behaviour, structure and evolution of the ISM. In turn, the evolution of SNRs is dependent on the environment in which they reside.

Here, we report on new radio-continuum and archival X-ray observations of \SNR, one of five SNRs in the LMC (B0540-693, N157B, B0532-710, DEM\,L241 and \SNR) with a known or candidate pulsar wind nebula (PWN) inside. \SNR\ was initially classified as an SNR based on the \einstein\ X-Ray survey by \citet[][source 1 in their catalogue]{1981ApJ...248..925L} and \citet[][their source 2]{1991ApJ...374..475W}. \citet{1983ApJS...51..345M} catalogued this object based on studies of optical and Molonglo Synthesis Telescope (MOST) survey data, and reported an optical size of $140\arcsec\times131\arcsec$. \SNR\ is also listed in the 408~MHz MC4 catalogue of \citet{1976AuJPA..40....1C} as a distinctive point-like radio source, for which \citet{1983ApJS...51..345M} later re-measured a flux density of 350~mJy. \citet{1984AuJPh..37..321M} detected this source with specific MOST pointings and reported a rather flat spectral index of \mbox{$\alpha$=--0.38.} \ROSAT\ detected X-ray emission from \SNR\ during the all-sky survey \citep{1998A&AS..127..119F} and in deeper pointed observations using the PSPC \citep[source 670 in the catalogue of][]{1999A&AS..139..277H} and HRI detector \citep[source 8 in][]{2000A&AS..143..391S}.

\citet{1998A&AS..130..421F} added further confirmation from radio data over a wide frequency range. \citet{1999ApJS..123..467W} classified it as an SNR with a ``diffuse face". \citet{1998ApJ...505..732H} presented X-ray spectra based on \asca\ observations. Their spectral modelling did not take into account the hard X-ray emission of the PWN, which was discovered later, and therefore overestimated its temperature. The most detailed study of \SNR\ was performed by \citet{2003ApJ...594L.111G} based on high resolution radio data obtained in 2002 with the Australia Telescope Compact Array (ATCA) and \cxo\ X-ray data from 2001. This lead to the discovery of the PWN inside the SNR and an age estimate of around 13\,000 years. \citet{2006ApJS..165..480B} reported no detection at far ultraviolet wavelengths based on data from the FUSE (Far Ultraviolet Spectroscopic Explorer) satellite. \citet{2006ApJ...652L..33W} detected this SNR in their Spitzer IR surveys as the object with the highest 70\,$\mu$m to 24~$\mu$m flux ratio of any SNR. However, due to high background emission at 70\,$\mu$m, in particular in the south western region of the remnant, they could only investigate the northern rim of the SNR shell. No radio pulsar within the area of \SNR\ was found in the systematic search within the LMC performed by \citet{2006ApJ...649..235M}. Also \citet{2008MNRAS.383.1175P} did not detect \SNR\ in their optical spectroscopic survey of LMC SNRs. Finally, \citet{2009ApJ...706L.106L} classified this LMC object based on its circular morphology as a core-collapse SNR. 

\section{Observations and data reduction}
 \label{section:observations}

\subsection{Radio-continuum}
 \label{datareduction_radio}

We observed \SNR\ with the ATCA on October 5, 1997, with the array configuration EW375 (ATCA Project C634), at wavelengths of 3~cm and 6~cm (8640~MHz and 4800~MHz). The observations were carried out in the ``snap-shot'' mode, totalling $\sim$1 hour of integration time over a 12 hour period. Source PKS B1934-638 was used for primary calibration and source PKS B0530-727 for secondary calibration. In addition to our own observations, we included unpublished 3~cm and 6~cm observations from project C1074 \citep{2003ApJ...594L.111G} and mosaic data taken from project C918 \citep{2010AJ....140.1511D}. When combined together, all 3~cm and 6~cm observations of \SNR\ total $\sim$11 hours of integration time. Baselines formed with the $6^\mathrm{th}$ ATCA antenna were excluded, as the other five antennas were arranged in a compact configuration. 

The \textsc{miriad} \citep{1995ASPC...77..433S} and \textsc{karma} \citep{1996ASPC..101...80G} software packages were used for reduction and analysis. More information on the observing procedure and other sources observed in this session can be found in \citet{2007MNRAS.378.1237B}, \citet{2008SerAJ.176...59C,2008SerAJ.177...61C,2010A&A...518A..35C}, \citet{2009SerAJ.179...55C}, and \citet{2010SerAJ.181...43B,2012MNRAS.420.2588B,2012RMxAA..48...41B}. Images were constructed using the \textsc{miriad} multi-frequency synthesis package  \citep{1994A&AS..108..585S}. Deconvolution was achieved with the {\sc clean} and {\sc restor} tasks with primary beam correction applied using the {\sc linmos} task. Similar procedures were used for the \textit{U} and \textit{Q} stokes parameters.

\begin{figure}[t]
 \begin{center}
 \resizebox{0.99\hsize}{!}{\includegraphics[clip=,angle=-90]{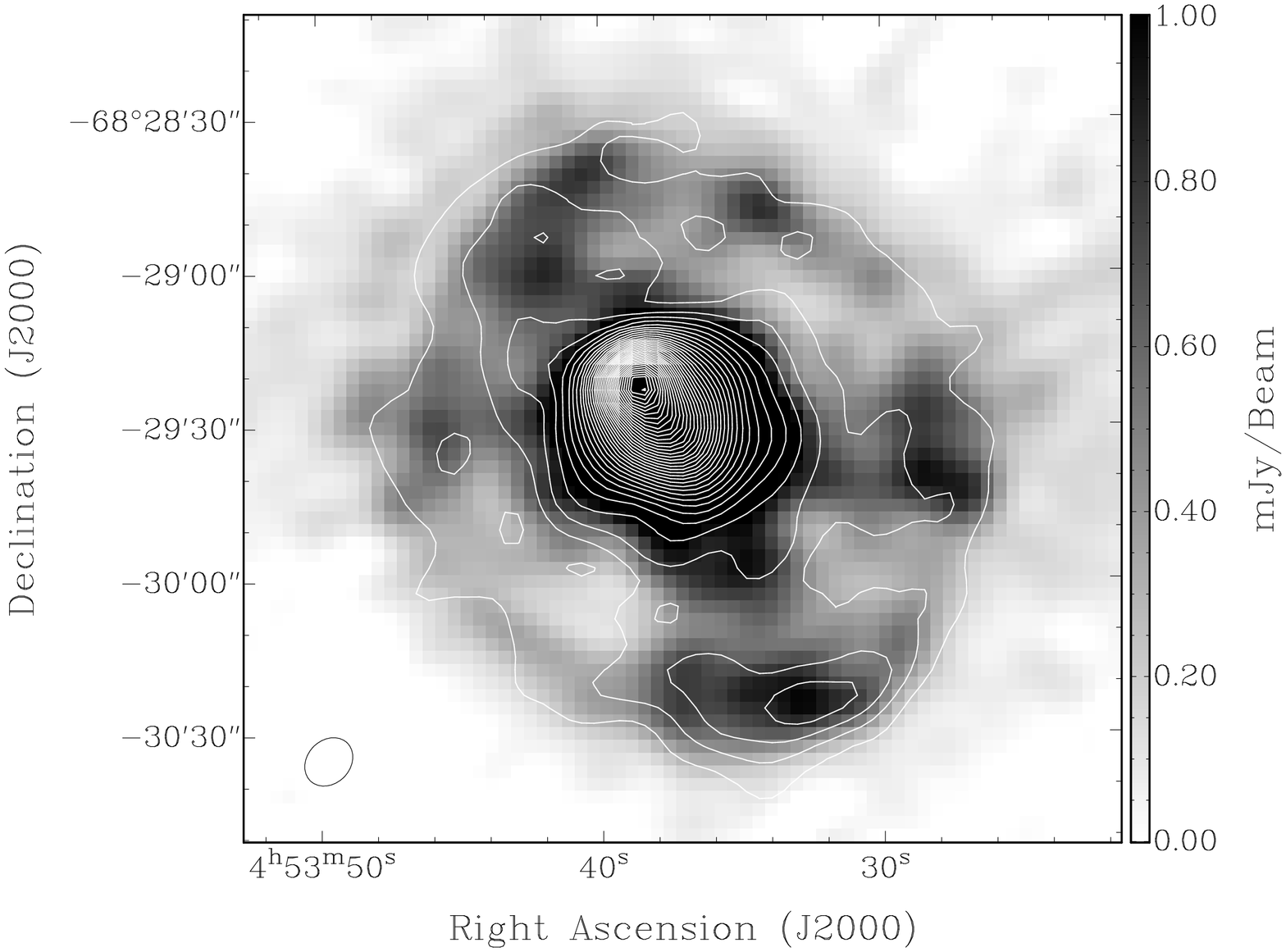}}
 \resizebox{0.99\hsize}{!}{\includegraphics[clip=,angle=-90]{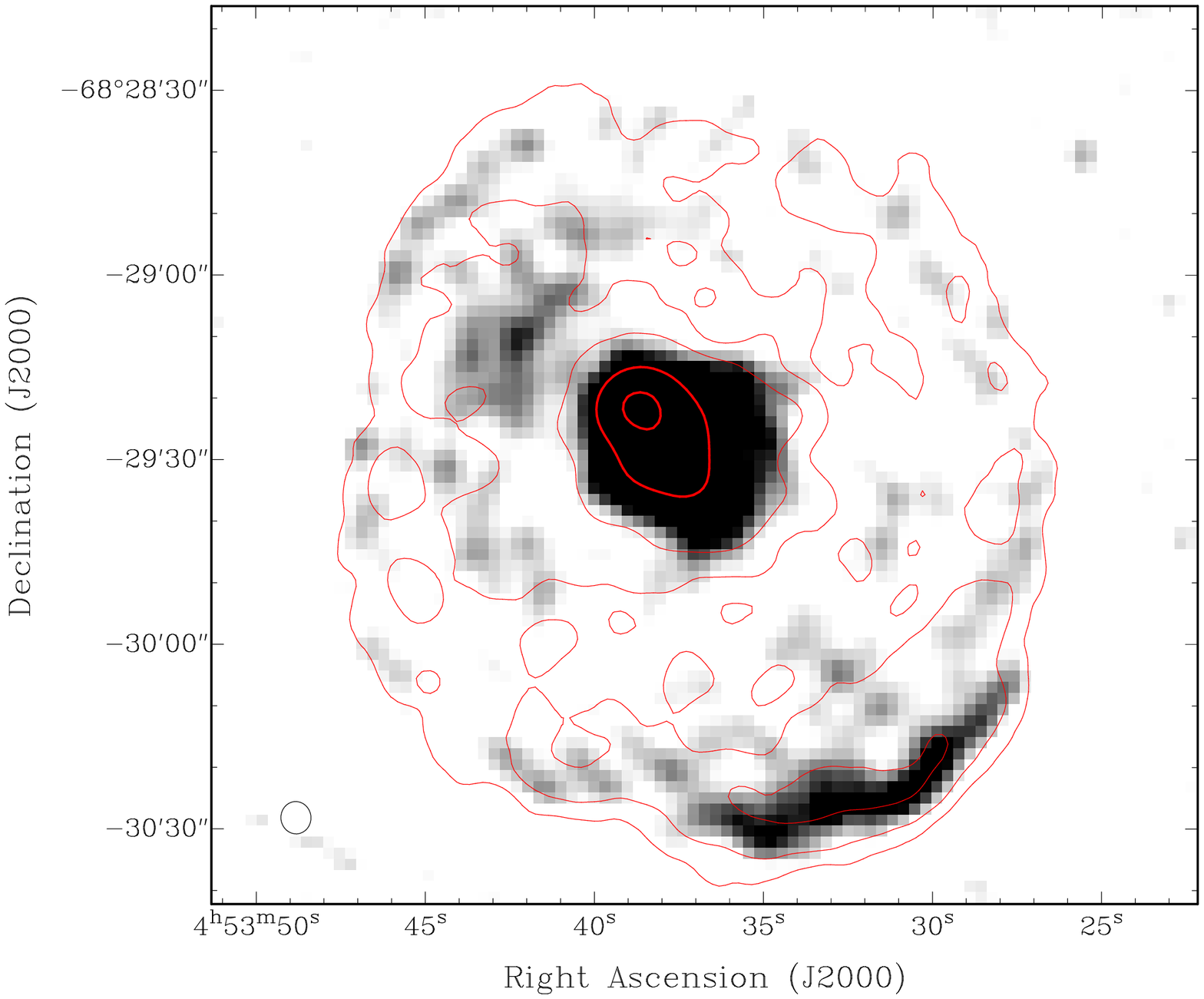}}
  \caption{ATCA observations of \SNR. 
          {\it Top:} 3~cm (8.6~GHz) image overlaid with 6~cm (4.8~GHz) contours. The contours are from 0.3 to 12.0~mJy/beam in steps of 0.3~mJy/beam with a beam size of 10.3\arcsec$\times$8.4\arcsec. 
          {\it Bottom:} 20~cm (1.4~GHz) image overlaid with 13~cm (2.4~GHz) contours. The contours are 0.12, 0.4, 1, 4 and 8 mJy/beam with the 20~cm image beam size of 5.2\arcsec$\times$4.9\arcsec. Beam sizes are indicated in the bottom left corners.}
  \label{fig-radio-ima}
 \end{center}
\end{figure}

\begin{figure}[h!]
\begin{center}
\resizebox{0.85\hsize}{!}{\includegraphics[clip=,angle=-90]{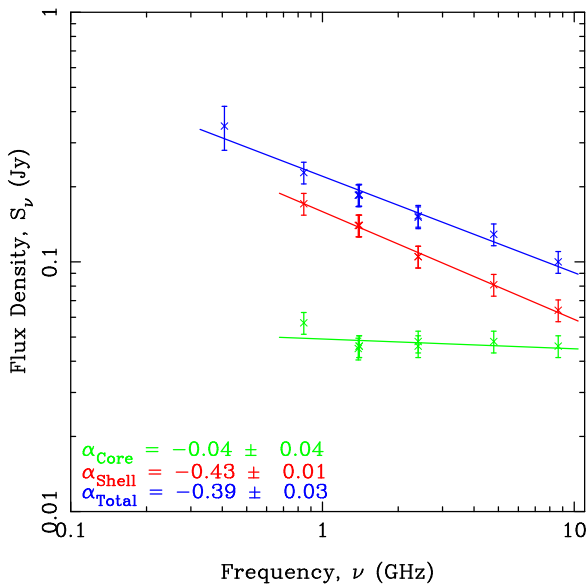}}
\caption{Radio-continuum spectrum of \SNR. The lines represent power-law fits to the
flux densities of the core (green), shell (red) and combined core and shell (blue).}
 \label{specidx}
\end{center}
\end{figure}

The ATCA 3~cm image is shown in Fig.~\ref{fig-radio-ima} together with 6~cm contours. We also reanalysed  the 20~cm and 13~cm observations from \citet[][ATCA Project C1074]{2003ApJ...594L.111G} and, while we achieved slightly better resolution and rms noise, the estimated flux densities stay essentially the same (Fig.~\ref{fig-radio-ima}). We measured the flux density of \SNR\ at six frequencies, which are summarised in Table~\ref{table-radiofluxes}. For the 36~cm (0.843~MHz, MOST) measurement we used the unpublished image described by \citet{1984AuJPh..37..321M} and for 20~cm (1.388~GHz, ATCA project C373) the image from \citet{2007MNRAS.382..543H}. We took the 20~cm (1.4~GHz) and 13~cm (2.4~GHz) measurements from \citet{2003ApJ...594L.111G} and added measurements at 20~cm (1.4~GHz) from the mosaics presented by \citet{2009IAUS..256...PDF-8} and \citet{2007MNRAS.382..543H}. 
The main source of uncertainty in the measured flux densities is the definition of the area best representing the SNR, which dominates the statistical error.
Based on different trials, we estimate that the combined flux density errors from all radio images used in this study \cite[apart from 73~cm where the error is in order of 20\%;][]{1983ApJS...51..345M,1976AuJPA..40....1C}, are $\sim$10\% at each given frequency.
Using the flux densities from Table~\ref{table-radiofluxes}, we estimated spectral indices for the PWN ($\alpha_{\rm Core}=-0.04\pm0.04$), the outer SNR shell excluding the core ($\alpha_{\rm Shell}=-0.43\pm0.01$), and the whole SNR ($\alpha_{\rm Total}=-0.39\pm0.03$); see Fig.~\ref{specidx}.

\begin{table*}
 \begin{center}
  \caption{Integrated Flux Density of \SNR}
 \label{table-radiofluxes}
  \begin{tabular}{cclcccccl}
\hline\hline\noalign{\smallskip}
\multicolumn{1}{c}{$\nu$} &
\multicolumn{1}{c}{$\lambda$} &
\multicolumn{1}{l}{ATCA} &
\multicolumn{1}{c}{R.M.S.} &
\multicolumn{1}{c}{Beam Size} &
\multicolumn{1}{c}{S$_\mathrm{Core}$} &
\multicolumn{1}{c}{S$_\mathrm{Shell}$} &
\multicolumn{1}{c}{S$_\mathrm{Total}$} &
\multicolumn{1}{l}{Reference} \\

\multicolumn{1}{c}{(GHz)} &
\multicolumn{1}{c}{(cm)} &
\multicolumn{1}{l}{Project} &
\multicolumn{1}{c}{(mJy)} &
\multicolumn{1}{c}{(\arcsec)} &
\multicolumn{1}{c}{(mJy)} &
\multicolumn{1}{c}{(mJy)} &
\multicolumn{1}{c}{(mJy)} &
\multicolumn{1}{c}{} \\
\noalign{\smallskip}\hline\noalign{\smallskip}
0.408 & 73 & MOST              & 50  & 157.3$\times$171.6 & ---& ---  & 350 & \citet{1983ApJS...51..345M}\\
0.843 & 36 & MOST              & 0.4 & 43                 & 57 & 171  & 228 & This work\\
1.388 & 20 & C373              & 1.2 & 40                 & 45 & 140  & 185 & This work -- MOSAIC\\
1.400 & 20 & C1074             & 0.1 & 7.3$\times$6.7     & 46 & 140  & 186 & \citet{2003ApJ...594L.111G}\\
1.400 & 20 & C1074, C373       & 0.05& 5.2$\times$4.9     & 46 & 140  & 185 & This work\\
2.400 & 13 & C1074             & 0.07& 9.2$\times$8.4     & 46 & 105  & 151 & \citet{2003ApJ...594L.111G}\\
2.400 & 13 & C1074             & 0.04& 7.8$\times$7.3     & 48 & 105  & 153 & This work\\
4.790 &\p06& C634, C918, C1074 & 0.1 & 10.3$\times$8.4    & 48 &  81  & 129 & This work\\
8.640 &\p03& C634, C918, C1074 & 0.1 & 10.3$\times$8.4    & 46 &  64  & 100 & This work\\
\hline
  \end{tabular}
 \end{center}
\end{table*}


\begin{figure}[h!]
\begin{center}
\resizebox{0.98\hsize}{!}{\includegraphics[angle=-90,clip=]{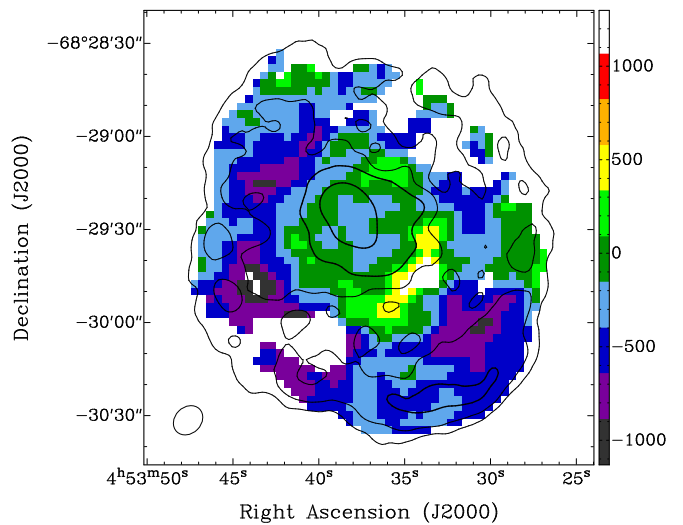}}
\caption{Spectral index map of \SNR\ using wavelengths of 3~cm, 6~cm, 13~cm, and 20~cm. The sidebar gives the spectral index scale, e.g. $-500$ corresponds to $\alpha=-0.5$. Contours (from 13~cm) are 0.12, 0.4, 1 and 4 mJy/beam. The beam size of  $10.3\arcsec \times8.4\arcsec$ is indicated in the bottom left corner.}
 \label{specmap}
\end{center}
\end{figure}

In Fig.~\ref{specmap} we show the spatial distribution of the spectral index for \SNR\ derived from wavelengths of 20~cm, 13~cm, 6~cm, and 3~cm. This image was formed by reprocessing all observations to a common $u-v$ range, and then fitting $S\propto\nu^\alpha$ pixel by pixel using all four images simultaneously. From this image, we can see that the vicinity around the central compact object exhibits relatively flat spectra with indices in the range --0.2$<\alpha<$+0.2, while the rest of the SNR exhibits spectra with $\alpha\sim-0.35$ to --0.7, typical for a standard SNR.

\begin{figure}[h!]
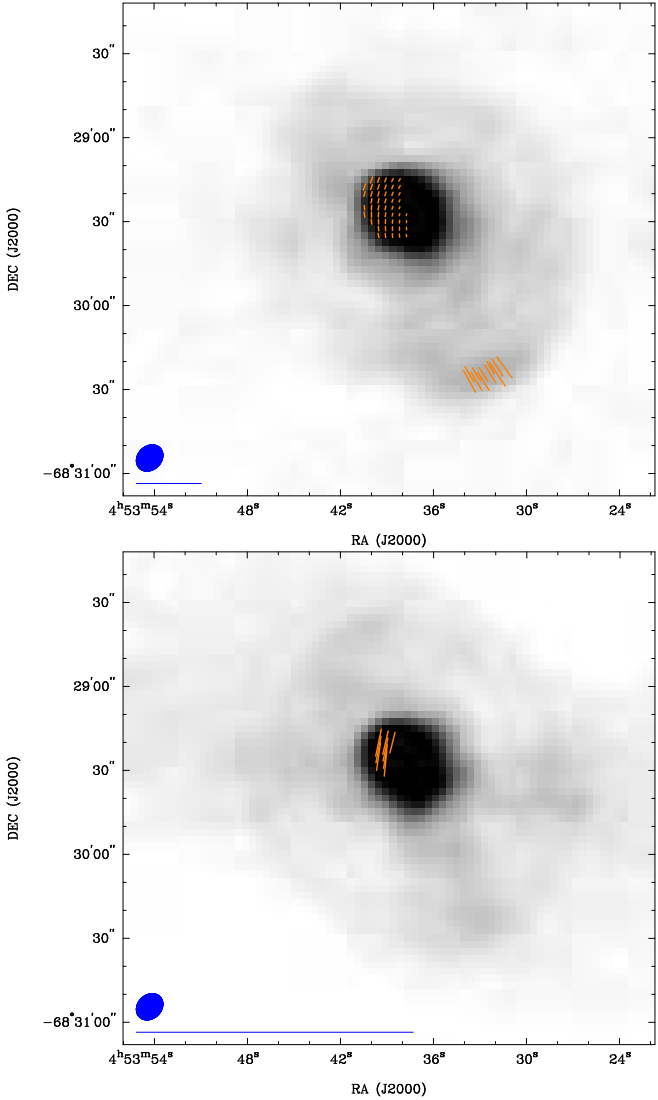

 \begin{center}
   \resizebox{0.95\hsize}{!}{\includegraphics[clip=,angle=-90]{6cmPOL.eps}}
   \resizebox{0.95\hsize}{!}{\includegraphics[clip=,angle=-90]{3cmPOL.eps}}
  \caption{Radio observations of \SNR\ at 6~cm ({\it top}) and 3~cm ({\it bottom}) matched to the 6~cm resolution. The blue circle in the lower left corner of the images represents the synthesised beam width of 10.3\arcsec$\times$8.4\arcsec. The length of the vectors represent the fractional polarised intensity at each position, and the orientation of the vectors indicates the mean position angle of the electric field (averaged over the observing bandwidth not corrected for Faraday rotation). The blue line below each circle represents the length of a  polarisation vector of 100\%.}
 \label{6cmpolar}
 \end{center}
\end{figure}

Linear polarisation images of \SNR\ at 6~cm and 3~cm were created using the \textit{Q} and \textit{U} Stokes parameters and are illustrated in Fig.~\ref{6cmpolar}. The fractional polarisation has been evaluated using the the standard \textsc{miriad} task \textrm{IMPOL}.  The majority of the polarised emission is seen in the core region. At 6~cm, the SNR core shows a mean fractional polarisation of $\sim7\%$ whilst along the shell (at the southern side of the SNR),  there is a pocket of quite strong uniform polarisation, with a mean value of $\sim32\%$ (Fig.~\ref{6cmpolar}, top) which corresponds to the observed total intensity brightening and possibly indicates varied dynamics along the shell. At 3~cm the mean fractional polarisation is $\sim9\%$.  Our estimated peak value is $P\cong$ 44$\pm12$\% at 6~cm and 12$\pm6$\% at 3~cm.

\subsection{X-rays}
 \label{datareduction_Xray}

\begin{figure}[t]
 \begin{center}
 \resizebox{0.98\hsize}{!}{\includegraphics[clip=]{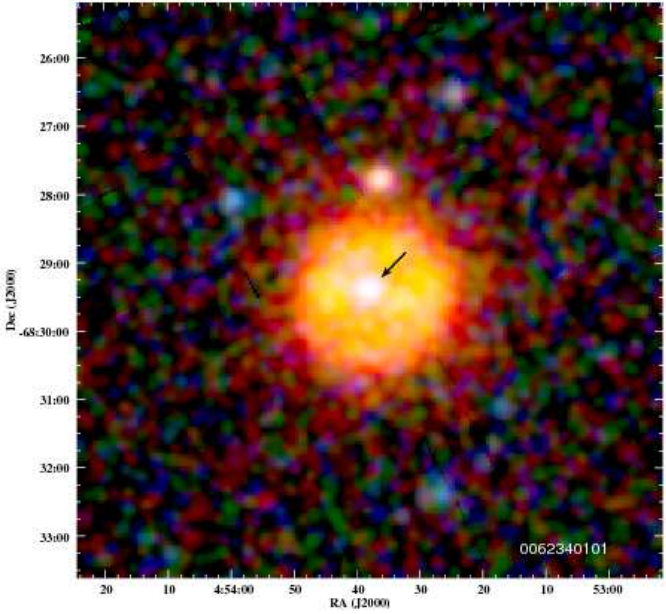}}
  \caption{Combined EPIC colour image of the field around \SNR. Red, green, and blue colours denote X-ray
           intensities in the 0.2$-$1.0, 1.0$-$2.0 and 2.0$-$4.5 keV bands. The strong soft X-ray emission from the SNR shell appears in yellow, while
           the bright white spot in the centre of the SNR (marked with a black arrow) is caused by the emission from the PWN. 
           Unrelated point sources are seen around the remnant.}
  \label{fig-ima-epic}
 \end{center}
\end{figure}

\SNR\ was observed with the instruments of the \xmm\ satellite. The observation (Obs. Id. 0062340101) was performed in March 2001 and here we present the results of the analysis of the X-ray data obtained by the European Photon Imaging Cameras (EPIC) and the Reflection Grating Spectrometers (RGS). The EPIC instruments were operated in full imaging modes using the medium optical blocking filters. Using the EPIC MOS1, MOS2 and pn CCDs yields sensitive X-ray observations of a $\sim$30\arcmin\ field in the 0.15-10 keV energy band. Technical descriptions of the EPIC cameras can be found  in \citet{2001A&A...365L..27T} and \citet{2001A&A...365L..18S}. The two RGS provide high-resolution X-ray spectra in the 0.3$-$2 keV band \citep{2001A&A...365L...7D}. A summary of the observations with the instrumental setups is given in Table~\ref{tab-xobs}. The \xmm\ data were analysed with SAS v10.0.0\footnote{Science Analysis Software (SAS): http://xmm.esac.esa.int/sas/.}. The data were affected by background flaring activity of moderate intensity. To produce images and spectra from the EPIC data we therefore removed intervals of high background while for the RGS the background level was below the recommended threshold (2 cts s$^{-1}$ in CCD number 9 of the instrument) and we used the full exposure time for spectral extraction. The final net exposure times used for spectral analysis are listed in Table~\ref{tab-xobs}. For the EPIC spectra we extracted events subject to the canonical set of valid pixel patterns (0 to 12 for MOS; 0 to 4 for pn) and rejecting events from pixels flagged as bad. \SNR\ is of roughly circular shape with a diameter of about 2\arcmin\ \citep{2003ApJ...594L.111G}, which causes a broadening of emission lines in the RGS spectra. This was taken into account when creating the response files with the task {\sc rgsrmfgen} by supplying the intensity profile of the remnant in the direction of the dispersion axis. This was created from an EPIC-pn image in the 0.2$-$2.0 keV band. A colour image produced from the combined EPIC data is shown in Fig.~\ref{fig-ima-epic}. The PWN at the centre of the SNR cannot be resolved by the \xmm\ instruments. Therefore, we extracted X-ray spectra from the full emission region of the remnant, including the PWN, and modelled the emission from the SNR and the PWN with different emission components.

\begin{table}
\caption[]{\xmm\ observation of \SNR\ on 2001-03-29.}
\begin{tabular}{lccr}
\hline\hline\noalign{\smallskip}
\multicolumn{1}{l}{Instrument$^{(a)}$} &
\multicolumn{1}{c}{Start time} &
\multicolumn{1}{c}{End time} &
\multicolumn{1}{c}{Exposure$^{(b)}$} \\

\multicolumn{1}{l}{configuration} &
\multicolumn{1}{c}{(UT)} &
\multicolumn{1}{c}{(UT)} &
\multicolumn{1}{c}{(s)} \\
\noalign{\smallskip}\hline\noalign{\smallskip}
EPIC MOS1 FF &  11:47:15 &  17:39:19 & 7366 \\
EPIC MOS2 FF &  11:47:15 &  17:39:19 & 7370 \\
EPIC pn eFF  &  13:00:34 &  17:40:02 & 3619 \\
RGS1 Spectro &  11:40:55 &  17:41:01 & 20820 \\
RGS2 Spectro &  11:40:55 &  17:41:01 & 20440 \\
\noalign{\smallskip}\hline\noalign{\smallskip}
\end{tabular}
\tablefoot{
$^{(a)}$ FF: full frame CCD readout mode with 2.6~s for MOS; EFF: pn `extended FF' with 200~ms frame time;
$^{(b)}$ Net exposure time used for spectral analysis after removing intervals of high background.
}
\label{tab-xobs}
\end{table}

All \xmm\ spectra were modelled simultaneously using {\sc XSPEC} \citep{1996ASPC..101...17A} version 12.7.0u, only allowing one relative normalisation factor for each instrument (using the EPIC pn spectrum as a reference with the factor fixed at 1.0). We fitted two-component models comprising of a thermal plasma emission and a power-law component, representing the thermal X-ray emission from the SNR and the harder, non-thermal emission from the PWN, respectively. For the plasma emission we used different models available in {\sc XSPEC} and report here on the results for the fits using either the plane-parallel shock (vpshock) model or the Sedov (vsedov) model \citep[][]{2001ApJ...548..820B}. For both cases we used NEI-version 2.0 and allowed some chemical elemental abundances to vary in the fit. For elements which do not contribute strongly to the X-ray spectra, the abundance was fixed to 0.5 solar, as is typical in the LMC \citep{1992ApJ...384..508R}. The solar abundance table was taken from \citet{2000ApJ...542..914W}. Both, thermal and non-thermal emission components in the model were attenuated by two absorption components along the line of sight, one accounting for the Galactic foreground absorption with a column density of 6\hcm{20} and solar abundances, and one for the absorption in the LMC with the column density as a free  parameter and abundances set to 0.5 solar for all elements heavier than He. For the absorption and emission components in the LMC a radial velocity of 278 \kms\ \citep[corresponding to a redshift of 9.27\expo{-4} ][]{1987A&A...171...33R} was adopted. The slope of the power-law was not well constrained in either model variant and we fixed the photon index at 2.0 \citep[typical values for PWN are between 1.5 and 2.2 ][]{2006csxs.book..279K}. For the vpshock model we also fixed the lower parameter $\tau_l$ of the ionisation time-scale range $\tau_l < \tau_u$  at 10$^8$ s cm$^{-3}$. The mean shock temperature and the electron temperature immediately behind the shock front can both be fitted by the Sedov model. First trial fits resulted in values fully consistent within their errors and in the following fits we forced the two temperatures to be the same. 

\begin{figure}[th]
 \begin{center}
 \resizebox{0.98\hsize}{!}{\includegraphics[angle=-90,clip=]{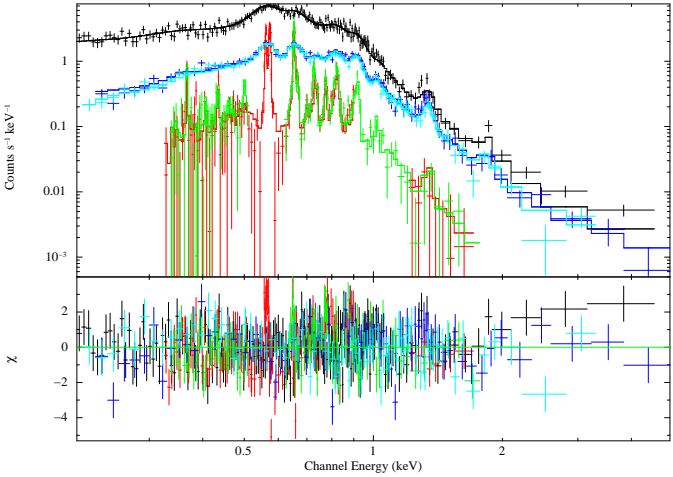}}
  \caption{Combined fit to the EPIC (pn: black, MOS1: blue, MOS2: light blue) and RGS (red: RGS1, green: RGS2) spectra of \SNR.
           The data are denoted by the crosses with error bars and the histograms represent the best-fit model. In the lower 
           panel the residuals are plotted in units of $\sigma$.
           The spectral extraction includes the emission from the SNR shell and the PWN and 
           the model consists of thermal plasma emission (Sedov model) from the shell and a power-law from the PWN.
           }
  \label{fig-xspec}
 \end{center}
\end{figure}

Both models have the same number of degrees of freedom (dof) and the Sedov model yields a formally better fit (see Table~\ref{tab-xspec}). The \xmm\ spectra, with the best-fit Sedov model, are shown in Fig.~\ref{fig-xspec}. The best fit parameters, which were free in the fit, are summarised in Table~\ref{tab-xspec} for both models. The total observed flux in the 0.2$-$10 keV band is 5.2\ergcm{-12} with a 12.0\% contribution of the power-law (derived from the EPIC-pn spectrum for the Sedov model). This converts to an absorption corrected luminosity of 7.0\ergs{37}.

\begin{table}
\caption[]{Spectral fits to the \xmm\ spectra of \SNR\ including its PWN.}
\begin{tabular}{lcc}
\hline\hline\noalign{\smallskip}
\multicolumn{3}{l}{Model vpshock ($\chi^2_{\rm r}$ = 1.48 for 574 dof)} \\
\multicolumn{1}{l}{component} &
\multicolumn{1}{c}{parameter} &
\multicolumn{1}{c}{value$^{(b)}$} \\
\noalign{\smallskip}\hline\noalign{\smallskip}
LMC absorption        & \nh\ (\ohcm{21})                 &  1.52 (1.39 - 1.69)    \\
shock temperature     & kT (eV)                          &   250 (234 - 269)      \\
abundance $^{(a)}$    & N                                & 0.052 (0.029 - 0.073)  \\
abundance             & O                                & 0.104 (0.094 - 0.113)  \\
abundance             & Ne                               &  0.17 (0.15-0.19)      \\
abundance             & Mg                               &  0.31 (0.25-0.36)      \\
abundance             & Si                               &  0.85 (0.54-1.18)      \\
abundance             & Fe                               &  0.20 (0.18-0.23)      \\
ionisation time scale & $\tau_u$ (10$^{11}$ s cm$^{-3}$) &  7.0  (5.0-9.0)        \\
\noalign{\smallskip}\hline\noalign{\smallskip}
\multicolumn{3}{l}{Model vsedov ($\chi^2_{\rm r}$ = 1.33 for 574 dof)} \\
\multicolumn{1}{l}{component} &
\multicolumn{1}{c}{parameter} &
\multicolumn{1}{c}{value} \\
\noalign{\smallskip}\hline\noalign{\smallskip}
LMC absorption        & \nh\ (\ohcm{21})               &  1.05 (0.98 - 1.17)    \\
shock/electron temp.  & kT (eV)                        &   168 (164 - 173)      \\
abundance             & N                              & 0.093 (0.057 - 0.133)  \\
abundance             & O                              & 0.20  (0.17 - 0.23)    \\
abundance             & Ne                             & 0.33  (0.28 - 0.39)    \\
abundance             & Mg                             & 0.52  (0.42 - 0.62)    \\
abundance             & Si                             & 0.81  (0.50 - 1.14)    \\
abundance             & Fe                             & 0.39  (0.34 - 0.44)    \\
ionisation time scale & $\tau$ (10$^{13}$ s cm$^{-3}$) & 2.4 ($>1.0$)    \\
\noalign{\smallskip}\hline\noalign{\smallskip}
\end{tabular}
\tablefoot{
$^{(a)}$ Elemental abundances are specified relative to solar \citep{2000ApJ...542..914W}.
$^{(b)}$ 90\% confidence ranges for one parameter of interest are given in parenthesis.
}
\label{tab-xspec}
\end{table}

\subsection{Optical}
 \label{datareduction_optical}

The Magellanic Cloud Emission Line Survey (MCELS)\footnote{MCELS : http://www.ctio.noao.edu/mcels/} was carried out at the 0.6~m University of Michigan/CTIO Curtis Schmidt telescope, equipped with a SITE $2048 \times 2048$\ CCD, which gave a field of 1.35\degr$\times$1.35\degr\ at a scale of 2.4\arcsec$\times$2.4\arcsec\,pixel$^{-1}$ \citep{2006NOAONL.85..6S}. Both the LMC and SMC were mapped in narrow bands corresponding to \Halpha, \OIII\ ($\lambda$=5007\,\AA), and \SII\ ($\lambda$ = 6716,\,6731\,\AA), plus matched red and green continuum bands. All the data have been continuum subtracted, flux-calibrated and assembled into mosaic images. The region around \SNR\ is shown in Fig.~\ref{mcels}) which demonstrates that [O\textsc{iii}] emission dominates around the outer edge of the SNR. We also note the enhanced 
H$_\alpha$ emission in the south-west direction adjacent to the SNR.

\section{Discussion}
 \label{discussion}
 
\begin{figure}[t]
\begin{center}
\resizebox{0.98\hsize}{!}{\includegraphics[angle=-90,clip=]{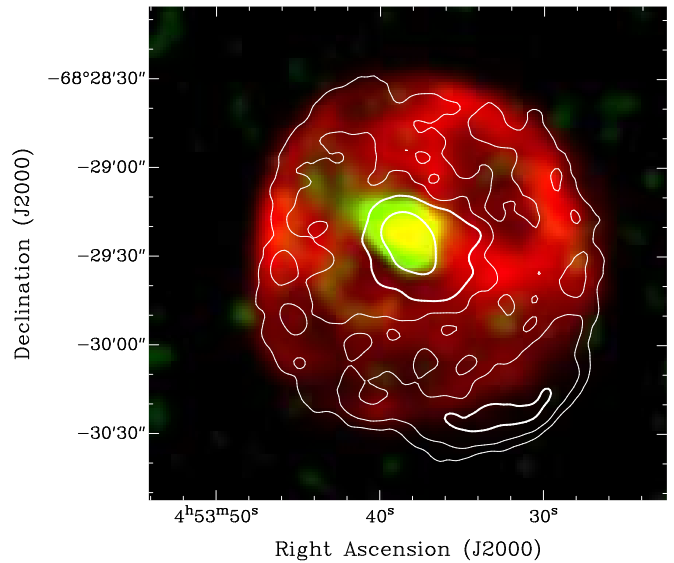}}
\caption{\cxo\ colour image of \SNR\ in the 0.3$-$2.0 keV (red) and 2.0$-$4.0 keV (green) energy bands overlaid with 13~cm contours. The contours are 0.12, 0.4, 1 and 4~mJy/beam and the \cxo\ image was smoothed to the same resolution as the radio image.}
\label{chandra}
\end{center}
\end{figure}

The \SNR\ in the LMC shows a prominent central core in radio and X-ray images, which is interpreted as a PWN \citep{2003ApJ...594L.111G}. The PWN is surrounded by a slightly elongated  shell centred at RA(J2000) = 4\rahour53\ramin37\fs2, DEC(J2000) = --68\degr29\arcmin30\arcsec\ for which we derive a  size of (128\arcsec$\pm$4\arcsec) $\times$ (120\arcsec$\pm$4\arcsec) (corresponding to (31$\pm$1~pc) $\times$ (29$\pm$1~pc) at a distance of 50~kpc) from the 13~cm radio data (obtained at position angles of 20\degr\ and 110\degr, running from north to east). The shell radio emission shows brightening along its southern rim. Our size estimate is in agreement with the diameter previously reported from radio data by \citet{2003ApJ...594L.111G}. From the intensity profile of the EPIC data we estimate a size of $\sim$140\arcsec\ (roughly in east - west direction, at 10\% of maximum intensity), which corresponds to $\sim$130\arcsec\ corrected for the angular resolution of the telescope. For a better estimate we re-analysed the \cxo\ data from 2001 presented by  \citep{2003ApJ...594L.111G}. From the \cxo\ image we measure 128\arcsec$\pm$4\arcsec (31$\pm$1~pc) as the largest extent, which is consistent with the radio extent (Fig.~\ref{chandra}). 
So in the following considerations we assume a radius of 15.5~pc.

\begin{figure}[h]
\begin{center}
\resizebox{0.98\hsize}{!}{\includegraphics[angle=-90,clip=]{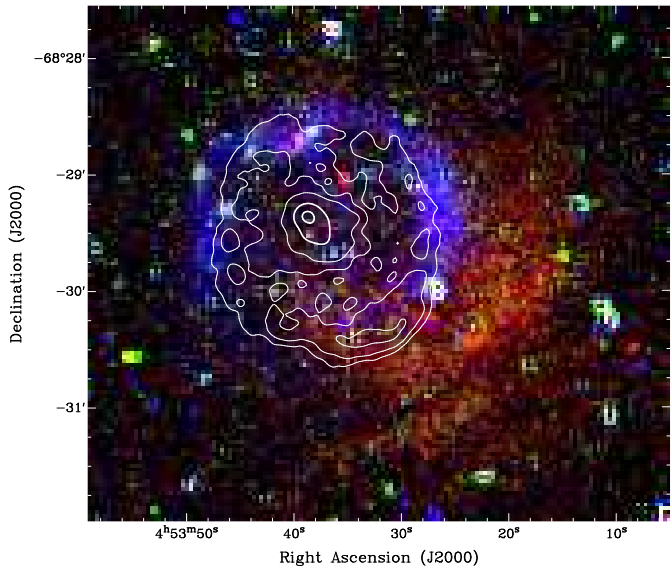}}
\caption{MCELS composite optical image \textrm{(RGB=H$_\alpha$,[S\textsc{ii}],[O\textsc{iii}])} of \SNR\ overlaid with 13~cm contours. 
  The contours are 0.12, 0.4, 1, 4 and 8~mJy/beam.}
\label{mcels}
\end{center}
\end{figure}

To estimate the age of the SNR from the vpshock model we used the relation t$_{\rm y}$\,=\,3.8\expo{2}\,R$_{\rm pc}$(kT)$_{\rm keV}^{-1/2}$ from \citet{2005ChJAA...5..165X} and assume the temperature derived from X-ray spectral modelling (0.25~keV) and the radius of the SNR, which results in $\sim$11.8~kyr, consistent with the age inferred by \citet{2003ApJ...594L.111G} from a preliminary X-ray spectral modelling. Also using a similar Sedov model and following \citet{2011A&A...530A.132O} who applied it to the spectrum of the SMC SNR IKT\,16, we derive -- with a shock temperature of 0.17 keV, and the normalisation of the Sedov model fit of 3.74\expo{-2} cm$^{-5}$ -- the following physical parameters for \SNR: electron density in the X-ray emitting material of 1.56\,cm$^{-3}$, dynamical age of 15.2~kyr, swept-up mass of 830\,\msun\ and an explosion energy of 7.6\expo{50}~erg. Our estimate for the density is higher than that derived by \citet{1998ApJ...505..732H} from \asca\ observations (without knowledge of the PWN contributing to the X-ray spectrum of the remnant) and consistent with the average value of the other LMC SNRs in that work. This directly reflects the higher density of the ISM in the LMC in comparison to the SMC \citep{2004A&A...421.1031V}. 

The shock temperature (and electron temperature in our Sedov modelling) is very low compared to other remnants in the MCs. The whole SMC sample modelled by \citet{2004A&A...421.1031V} and \citet{2008A&A...485...63F} in a similar way exhibits temperatures between 0.26~keV and 1.8~keV. The low temperature leads to a relatively high dynamical age for \SNR. The high swept-up mass is also consistent with an SNR well evolved into its Sedov phase. As discussed in \citet{2004A&A...421.1031V} the higher ISM density and higher abundances in the LMC compared to the SMC can lead to a faster evolution to the radiative cooling stage.

The abundances derived from our spectral analysis exhibit some model dependences, but are generally consistent or somewhat lower than average abundances in the ISM of the LMC. The only exception is Si which is overabundant with respect to the other elements. This suggests that the X-ray spectrum is dominated by emission from swept-up ISM material, which makes it difficult to draw conclusions on the type of supernova explosion from abundance measurements. In the case of \SNR\ the association of the PWN with the remnant and its morphology favour a core-collapse supernova \citep{2009ApJ...706L.106L}.

The radio spectral index of $\alpha_{\rm Shell}=-0.43$, confirms the non-thermal nature of the SNR shell emission in the radio band, while the flat spectral index for the core  of $\alpha_{\rm Core}=-0.04$ is typical for a PWN. The overall spectral index of $\alpha=-0.39$ is slightly ``flatter'' in comparison with typical values of $-0.5$ for SNRs \citep{1985ApJS...58..197M,1998A&AS..130..421F}. The radio spectra definitely confirm the PWN nature of the central object. At higher frequencies (Fig.~\ref{specidx}) the flux density decreases as expected whereas at lower frequencies non-thermal radiation of the shell dominates. Overall, the radio-continuum properties of \SNR\ are very similar to SNR~B0540-693, the most prominent SNR with a PWN in the LMC \citep{1993ApJ...411..756M}. 

\citet{1983ApJS...51..345M} found distinctive optical connections to X-ray and radio features of this SNR. As can be seen in the MCELS images (Fig.~\ref{mcels}), the \OIII\ emission dominates in the outer parts of the remnant. This could indicate an oxygen-rich type of SNR and suggests a type~Ib SN event -- the explosion of a massive O, B, or WR star \citep{2005MNRAS.360...76A}. We also note that prominent H$\alpha$ emission is visible in the south-west, 
which might suggest that higher density ISM material is causing the brightening in the radio shell emission at the outer rim of the SNR in this direction.
While no molecular clouds are reported in this region \citep{2008ApJS..178...56F,2009ApJS..184....1K}, we do note that \SNR\ lies in a region of low \ion{H}{i} column density with the rim density highest in the south west \citep{1998ApJ...503..674K,1999AJ....118.2797K}. The radio spectral index (Fig.~\ref{specmap}) in this south-west region indicates non-thermal emission $(\alpha \sim -0.4)$. Careful comparison (after correcting for the rotated presentation) of the 24 \& 70 $\mu$m infra-red data shows no correspondence between the enhanced background reported by \citet{2006ApJ...652L..33W} and \SNR. In contrast however, the \cxo\ image reveals only weak X-ray emission at the location of the bright radio feature. This could be explained by the higher density material being located in front of the rim of the SNR, suppressing the X-rays, or, together with the high polarisation of the 6\,cm radio emission in this area, might favour a locally increased magnetic field strength enhancing the non-thermal radio emission. Future spatially resolved X-ray spectroscopy should be able to resolve this question.

%

\begin{acknowledgements}
This publication is partly based on observations with \xmm, an ESA Science Mission with instruments and contributions directly funded by ESA Member states and the USA (NASA).
The \xmm\ project is supported by the Bundesministerium f{\"u}r Wirtschaft und Technologie/Deutsches Zentrum f{\"u}r Luft- und Raumfahrt (BMWI/DLR, FKZ 50 OX 0001) and the Max-Planck Society.
We used the {\sc karma} software package developed by the ATNF. The Australia Telescope Compact Array is part of the Australia Telescope which is funded by the Commonwealth of Australia for operation as a National Facility managed by CSIRO. 
The Magellanic Clouds Emission Line Survey (MCELS) data are provided by R.~C. Smith, P.~F. Winkler, and S.~D. Points. The MCELS project has been supported in part by NSF grants AST-9540747 and AST-0307613, and through the generous support of the Dean B. McLaughlin Fund at the University of Michigan, a bequest from the family of Dr. Dean B. McLaughlin in memory of his lasting impact on Astronomy. The National Optical Astronomy Observatory is operated by the Association of Universities for Research in Astronomy Inc. (AURA), under a cooperative agreement with the National Science Foundation.
\end{acknowledgements}

\bibliographystyle{aa}
\bibliography{MC}

\end{document}